\documentstyle[11pt,newpasp,twoside,epsf]{article}
\def\edcomment#1{\iffalse\marginpar{\raggedright\sl#1\/}\else\relax\fi}
\reversemarginpar

\begin{document}
\title{Modern Techniques in Galaxy Kinematics:\\ 
CDI and the Planetary Nebula Spectrograph}
% \author{N. G. Douglas            %  \& Members of the PN.S team$^1$}
% \affil{Kapteyn Institute, Groningen, Netherlands,
%  ndouglas@astro.rug.nl}

% \author{

\author{N. G. Douglas,
K. Kuijken, A.J. Romanowsky,
M.R. Merrifield, \\
M. Arnaboldi, K. Freeman, K. Taylor}
\affil{Members of the PN.Spectrograph Team}

\begin{abstract}
	We report here the successful
	commissioning of the  \\
        PN.Spectrograph,
	the first special-purpose instrument for the \\
	measurement
	of galaxy kinematics through the PN population.
\end{abstract}

\section{Introduction}

Planetary nebulae: to most of the participants at this conference, 
beautiful \\
objects displaying a range of structures,
spherical, bipolar and
rhomboidal,\\ riotous colours, and the
complex spectral signature of ionised gases, shocks, and dust. To others,
the extragalactic crowd, they are seen as featureless
points of light with the simplest of all possible spectra, a solitary
green emission line. 

\section{Observations of PNe in external galaxies}

The peculiar obervational properties of PNe at large distance have made
them popular probes of galaxy dynamics.  A large fraction of the total
luminosity of the central star is re-emitted at the 5007\AA\ [O{\sc
III}] wavelength, making PNe easy to detect as individual objects even
at the distances required by curent galaxy studies ($\sim 20$-30~Mpc). 
Detection is usually limited by the galaxy's own diffuse light, so PNe
are most useful in the fainter, outer parts.  Therefore they complement
the use of integrated stellar spectroscopy, which is generally limited
to the central 1-2 effective radii.  Determining the radial velocities
of the PNe is in principle straightforward on account of their narrow
emission-line spectra. 

At IAU Symposium 180 we reported plans for
a new instrument, designed to measure the positions and radial velocities
of PNe in a single observation. We are pleased to report
that this ``Planetary Nebula
Spectrograph" was commissioned at the William Herschel Telescope 
(the 4.2m telescope at
La Palma Observatory) on July 16 2001, just a few months before the current 
meeting. 

\section{Why a dedicated instrument?}

Detection of PNe in external galaxies is normally done by means of an 
on-band/off-band survey to detect emission-line objects at the [O{\sc III}] wavelength,
followed by multi-object spectroscopy to measure the radial velocities. These
two steps are usually completely disjoint in terms of the instrument and
telescope used, leading to significant astrometric errors so that
in practice the spectroscopy often returns unsatisfactory
results in terms of completeness and accuracy. 

The PN.Spectrograph
 uses slitless spectroscopy both to find the PNe and to
measure their radial velocities. By
splitting the light from the telescope between two gratings, a pair of
images is obtained in  which the dispersion directions are opposite: we call
this {\em counter-dispersed imaging} (CDI). From these two frames, 
the velocities of the point-like objects can be obtained.

\begin{figure}
\plottwo{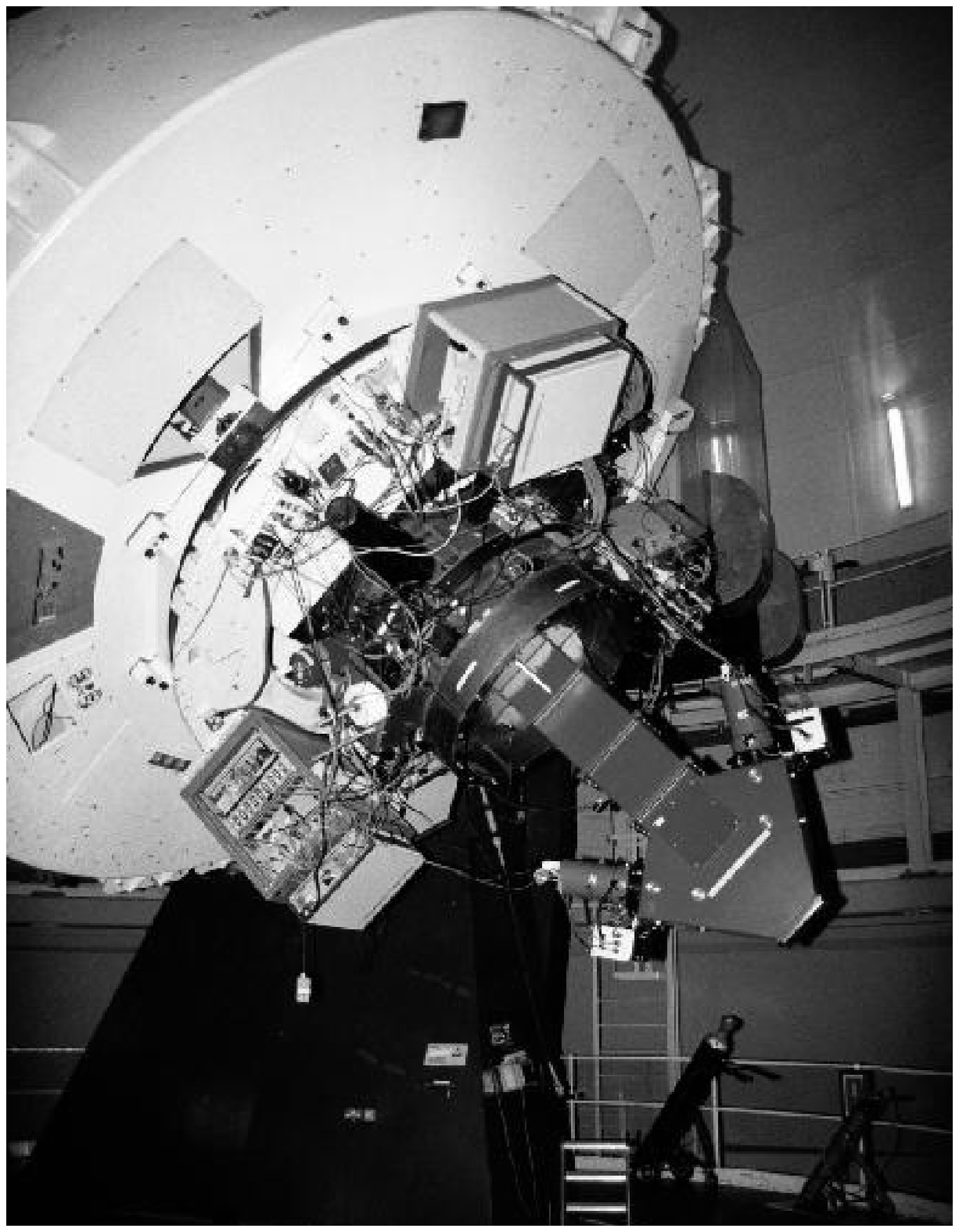}{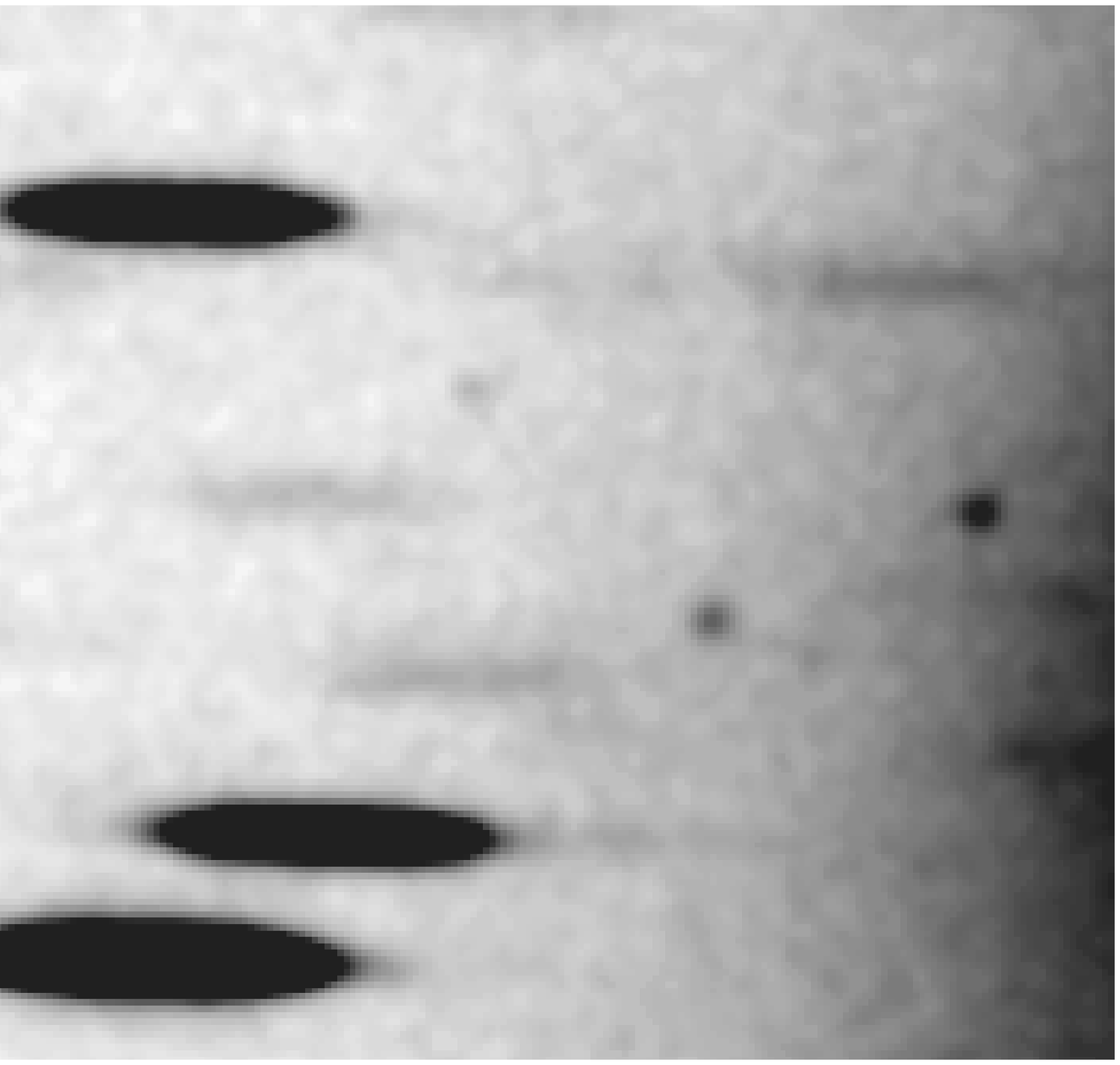}

\caption{Left: the PN.S at the Cassegrain focus of the 
WHT - visible in the picture are the two cryostats whose axes
point to the dual grating assembly at which the pupil is split between the
two cameras. Right: a small section of a typical image, showing
several PNe (points) as well as
three bright and several fainter stars (extended)}
\end{figure}

Thus, the entire observation consists of a single deep integration
with the PN.Spectrograph. The total time spent is longer than that required
by the traditional  on-band/off-band survey, but not by much. For one thing,
switching between filters is no longer required, since 
emission-line sources can be distinguished on the basis of their shape 
(see Fig.~1). Also, 
the dedicated design has allowed us to optimise all the components
at 5007\AA\ and thus reach high optical efficiency (33\%). More
importantly, the
entire multi-object spectroscopy stage is rendered redundant. The fact that
(relative) velocities are determined by the displacement of point-like
objects on the stable medium of a CCD lends itself to very high accuracy.
The absolute velocity scale is fixed by use of a calibration mask
which, like the field of interest, is registered simultaneously
in both arms of the spectrograph.
The PN.Spectrograph is currently returning exciting data and progress may be
monitored through our 
website\footnote{PN.S website: { http://www.astro.rug.nl/{\small $\sim$}pns}}.

\end{document}